# Оглядові та проблемні статті
# Reviews and topical articles

О.В.Градов
Ф.А.Насиров
А.А.Скрынник
А.Г.Яблоков

Институт энергетических проблем химической физики РАН им. В.Л.Тальрозе Лаборатория биологического воздействия наноструктур (005) Москва, Российская Федерация

УДК: 57.085.2+576.08+577.12+611.018.8+541.18+531.715.1

## ПРОСТОЕ МЕТОДИЧЕСКОЕ ПРИСПОСОБЛЕНИЕ ДЛЯ МИКРОИНЪЕКТОРНЫХ МАНИПУЛЯЦИЙ И ИЗМЕРЕНИЙ НА ЭЛЕКТРОМОРФОЛОГИЧЕСКОМ ЧИПЕ ПРИ МИКРОИНТЕРФЕРОМЕТРИЧЕСКОМ КОНТРОЛЕ ИНТЕРФЕЙСНЫХ И МЕМБРАННЫХ ПРОЦЕССОВ НА ДИАПАЗОНЕ ТОЛЩИН ОТ 50 ДО 10000 АНГСТРЕМ ПОД РАЗНЫМИ УГЛАМИ

**Реферат.** Микроманипуляции, перфузии и измерения, проводимые с использованием стеклянных микроэлектродов, являются классической техникой экспериментально-морфологических и мембранно-электрофизиологических исследований на уровне отдельных клеток и мембранных поверхностей. Стандартный (эффективный) диаметр стеклянного микроэлектрода в оконечной области составляет от 500 нм до менее, чем 100 нм, что препятствует использованию стандартных оптических микроскопов для его наблюдения. Нами предлагается конфигурация установки для микроперфузии, микроманипуляции и микроэлектродных измерений с интерферометрическим контролем процессов на интерфейсе клеток и капилляров/электродов.

**Gradov O.V., Nasirov F.A., Skrynnik A.A., Jablokov A.G. A simple device for microinjections, manipulations and measurements using an electromorphological chip under microinterferometric control of the interface and membrane processes at the thickness range of 5-1000 nm at different angles.**
**ABSTRACT.** Micromanipulations, perfusions and measurements performed using glass microelectrodes filled with an electrolyte is a conventional technique for experimental morphological and membrane electrophysiological studies at a single cell and membrane surface level. The typical (effective) diameter of the end of the glass microelectrode is from 500 up to less than 100 nm, which prevents one from observing it using a standard optical microscope in accordance with the optical resolution criteria, since the diameter less than 500 nm is indistinguishable within the interference zone. Microprocessor programming of the puller (microforge) that provides pulling and tearing allows to obtain in certain regimes the adjusted diameter and shape of the micropipette tip, although this result is not fully controlled due to the above limitations. In this connection it is necessary to design the control devices for the micropipette tips both at the preparation and operation stages (intracellular or extracellular insertion). This method also should provide visualization of the processes occurring upon interaction of the microelectrode tip with the cell in real time, depending on the electrode type and state, which allows to level the artifacts arising with the systematic error frequency from the uncontrolled operation of the micropipette tip after different ways of the microelectrode filling with the electrolyte. We propose an installation scheme that solves the above problems by means of introducing an interferometric device for microscopic control of the microelectrode and micromanipulator or microperfusor, for the first time for a given type of optical instruments combined with the interferometric optical scheme.

**Введение**

Проблема взаимодействия между микроэлектродом и клеткой в тонком эксперименте, с позиций биофизики поверхности [1-3], является одновременно проблемой физической химии поверхности электрода и биофизической химии интерфейсных с ним областей клетки, завися одновременно от электростатических характеристик конкретных марок специального стекла и конкретного электролита в стеклянном микроэлектроде по физическим источникам сигнала зондирования, и от биологической геометрии эксперимента (внеклеточное, внутриклеточное или интерфейсное отведение конкретного электрода) по ожидаемому характеру его сигналов. При этом, в отсутствие точных и достоверных техник наблюдения кончика микроэлектрода в среде (так как при диаметре конуса менее 500 нм он теряется в интерференционной кайме на обычной оптической цифровой или аналоговой микрофотографии) ничего конкретного о том или ином типе процессов в активной зоне интерфейса клетка-электрод (на острие последнего) в экспериментальных статьях определить невозможно. Работы, выполняемые на стандартных световых микроскопах, физически лишены данных о процессах на границе «электрод-клетка» и «электрод-среда» в силу ограниченности их разрешения критерием Рэлея, а продвижение в ультрафиолетовую область изменяет сам ход подобных процессов, негативно воздействует на клетку и, как следствие, не обеспечивает неразрушающий контроль в ходе эксперимента. Так как явления диффузии, ионофореза, электроосмоса и сопутствующие им флюидные эффекты работают на регистрируемых оптическими средствами пространственных масштабах, однако только на масштабах, фиксируемых интерферометрическими методами, логично опробовать, предварительно теоретически обосновав, интерференционные методы контроля «в реальном времени», *in situ*, в процессе эксперимента.

Физические экспериментальные интерферометрические исследования диффузии имеют началом 1940-е годы, работы Лонгсворта, Гостинга и Лайонса с соавторами [4-11]. Эти исследования, базировавшиеся, зачастую [9, 10], на методе Гуи, генеральная физическая теория которого была разработана Гостингом и Онзагером [12], составили основу интерферометрии в физической химии диффузии и биохимии (например – интерферометрические измерения для анализа диффузии глюкозы, мочевины, глицина и т.д. [7, 9, 10]) соответствующих процессов как *in vitro*, так и *in vivo*. Альтернативные работы, выполненные позже по схеме интерферометра Жамена (либо, что точнее, интерферометрического рефрактометра Жамена) [13, 14], приводили к аналогичным результатам и также способствовали развитию интерференционных техник в биохимии, легших в основу некоторых типов хроматографических детекторов, измеряющих в качестве первичного сигнала диффузию. Интерференционные диффузионные измерения для нужд хроматографии актуальны до настоящего времени, так как цеолиты, использующиеся в большом количестве методов как набивка колонок, эффективно контролируются в динамике интерференционными методами, особо микроинтерферометрического плана [15-21], что свойственно для большинства исследований диффузии, сопряженных с кристаллическим состоянием вещества [22-24]. Однако клетка не является, в качественном смысле термина, ни твердым телом, ни жидкостью – к ней неприменимы в чистом виде ни жидкофазные техники, которые адекватны для wet biochemistry [7, 9, 10] или аналогичных измерений небиологических жидкостей [25] (тот же метод Гуи), ни твердофазные подходы, включая использующиеся для измерений при процессах затвердевания типа градиентной солидификации [26]. Более того, по принципам биофизической термодинамики, невозможно сравнивать гомогенную диффузию в безмембранной стремящейся к равновесию системе (например – диффузию солей некоторых органических кислот в воде, хорошо визуализируемую стандартным интерферометрическим путём, например – с использованием интерференционного метода Гуи [27]), и в гетерогенной / структурированной или частично упорядоченной (soft matter) среде, к типу которых относят, в частности, многофазные смеси биополимеров и цитоплазму. Следует учитывать подвижные границы молекулярной сети [28] и свойства самих полимеров [29,30] клетки, рассматривая её в контексте физической химии как гельсодержащую структуру и структуру, способную к золь-гель-переходам (с первой половины XX века [31-34] до настоящего времени [35-40]) и их регуляции внешними физическими (электрическими, термическими, фотофизическими, радиационными) факторами, так как при данном рассмотрении приобретает «эвристическую ценность» применение к клеточным объектам диффузиометрических интерферометрических методов, адекватных для гелевых белковых систем [41].

Существенную трудность представляет то, что интерфейс цитоплазмы клетки и среды в диффузионном аспекте опосредован мембраной, в то время как возможности анализировать эффлюкс с обратной стороны мембраны / изнутри клетки интерферометрическими методами не представляется возможным. Однако эти методы эффективно используются для измерений и визуализации:

1) поверхностной мембранной топографии [42];

2) водной проницаемости и осмотических явлений в мембранных структурах [43];

3) фазовых переходов в биологических мембранах [44];

4) реологических свойств, растяжимости и деформаций биологических мембран [45,46];

5) ориентационных эффектов в липидных мембранах на чипе [47] (в том числе – векторной ориентации мембранных белков при использовании рентгеновской интерферометрии и голографии [48] или векторной ориентации ионных каналов по данным корпускулярной – нейтронной интерферометрии [49]);

6) адгезии клеток к поверхности [50];

7) эффектов взаимодействия компонентов мембран с разными лигандами (в таких случаях используют, как правило, интерферометрию обратного рассеяния [51,52], даже когда речь в работах идёт о мембранных белках [53]; анализ взаимодействия с пептидами зачастую производится поляризационно-интерференционными методами [54]);

8) коллективных движений в мембране различной амплитуды [55];

9) функционально-морфометрических карт мембран с использованием оптики Номарского [56] и по данным самоинтерференции флуоресценции [57];

10) активности пролиферации клетки – с позиций анализа ILMS (Internal Limiting Membrane Specimens) [58];

11) нейрофизиологических внутриклеточных и мембранных процессов на уровне нейронов в развитии и реакции, и их формирующегося в результате этой активности коннектома [59].

Исходя из пунктов приведенного списка, связанных с межфазным взаимодействием или взаимодействием мембранных компонент с внешними агентами, рационально предположить, что нельзя рассматривать клеточную мембрану как интактную или инертную структуру, как это принято в простой механической трактовке, порождающей упрощенный подход к данным биомембранной интерферометрии (формально апеллирующий к целям обычной инженерно-физической интерферометрии оболочек и деформирующихся поверхностей [60-64], что в принципе верно, но может быть применено более к «анатомическим» макромасштабным или мезомасштабным «мембранам» – хориоаллантоисной оболочке [65], тимпанальной [66-69], базилярной [70], fenestra cochleae / fenestra rotunda [71], эндотелия роговицы [72] и т.п.). При иных подходах к интерпретации – изменяется спектр явлений, которые могут быть исследованы по принципам микроинтерферометрии в реальном времени. Чем большее число динамических в основе своей явлений исследователь сможет проконтролировать в ходе микроэлектродного и микроперфузионного (требующего, как минимум, ещё учета текучести мембран и жидкостей, вводимых в них [73]) эксперимента, тем более он будет застрахован от артефактов и тем более точным будет представление о комплексных изучаемых им механизмах, происходящих в ему подконтрольной экспериментальной среде.

Если не вводить чисто твердотельные интерференционные диффузионные ограничения [74], то измерять диффузию через ограничивающие и прерывающие её системы типа мембран и границ раздела, а также пространственных барьеров прерывания, возможно, даже исходя из старых подходов [75], также как и диффузию через абстрактные реактивные оболочки (ESPI –спекл-интерферометрией [76] и другими методами), по интерференционным полосам. Первые результаты по определению коэффициентов диффузии интерпретацией интерференционных полос в физической химии относятся к границе первой и второй половин XX века, а подходы к скоростному расчету коэффициентов диффузии по рэлеевским полосам внедрены в 1960-е годы [77] (эти работы были продолжены впоследствии – см. статьи по анализу коэффициентов диффузии по прогрессии интерференционных полос 1980-х гг. [78] и наиболее современные в инструментальном плане публикации по определению коэффициентов диффузии по полосам регистрограмм интерферометров Маха-Цендера, являющихся модификацией интерферометра Жамена, упомянутого выше [79]). Однако специализированных систем анализа реактивности мембран клеток (по диффузии в процессе эксперимента) с интерференционными полосами до настоящего времени не создано. Можно было бы сопрячь это с отсутствием моделей, методов и идеологий расшифровки для сложных диффузионных режимов, в частности, сопряженных (в случае цитоэлектрофизиологически активных мембран) с биоэлектрогенезом. Однако: для задач мембранного электроосмоса на границе раздела фаз существуют развитые модели [80], а для рассматриваемого как функционально мембраномиметический процесс обратного осмоса существует развитая индустрия микроинтерферометрического, в частности голографического интерферометрического контроля процессов [81-86]. Следовательно, объективных (т.е. физико-химических) предпосылок для отсутствия техники интерферометрического контроля диффузии, ионофореза, электроосмоса и сопутствующих флюидных эффектов в пограничном слое микроэлектрода и клетки при экспериментально-биологических работах, в принципе, не имеется. Ниже описывается создание простейшего DIY-стенда микроинтерферометрических измерений вблизи микроэлектрода, обеспечивающего контроль ряда явлений массопереноса в зоне интерфейсов клетки, среды и электрода.

**Конструкция устройства**.

Нами в качестве исходного интерферометрического модуля использовался стандартный промышленный многолучевой микроинтерферометр МИИ-11 индустриального назначения, в котором многолучевая интерференция имплементируется посредством подборки стеклянных пластин с избирательным (селективным) отражением излучения. Эффективность применения данного аппарата, в идеальной конфигурации, достигается при использовании когерентного – лазерного монохроматического источника с заданной длиной волны, так как при наложении пластины на объект или ином контакте исследуемого объекта с ней при монохроматическом облучении в промежуточном пространстве наблюдается многократное отражение излучения, фиксируемое цифровой камерой на окуляре прибора. Использование камеры вместо окуляр-микрометра позволяет выполнять на ЭВМ измерения интерференционных полос, которые по механизму генерации и локализации соответствуют особенностям структуры микрообъекта. Точность подобных измерений чрезвычайно высока. Реализована возможность измерений по толщине от 50 до $100^3$ Å. Это адекватно толщинам биологических мембранных структур как с учетом кортикальных и экстрацеллюлярных прилежащих областей, так и без учета таковых. Нами был использован микроинтерферометр МИИ-11 исходно с некогерентным облучением, однако затем были установлены три лазерных диода (405 нм, 532 нм, 650 нм). Благодаря этим элементарным усовершенствованиям, была реализована возможность спектрозонального (т.е. в R-, G-, B- фотоколориметрических диапазонах спектра) измерения разных структур. Нами данные источники использовались только в монохроматическом режиме без перекрытия, так как в противном случае суммарная энергия выводила из строя объект/или матрицу. Объектив был подобран специально. Окуляр-микрометр, работавший на некогерентной версии, заменен цифровой камерой с соответствующим расширенным программным обеспечением (с ImageJ). В ряде электрофизических и флюидных экспериментов стеклянные пластины с селективным отражением совмещались с чипом. По данным методических материалов ИТМО, для режима многолучевой интерференции возможно расширение нижнего предела измерений до $h_1=0{,}003$ мкм или неоднородности от $R_a$ 0,16 до $R_a$ 0,008. Общий вид установки показан на рис. 1.

В качестве микроинъектора / микроманипулятора / микроперфузора использовался для первых экспериментов инъектор стереотактической установки с граммометром, ненужным на практике при использовании под углом без контакта с твердой поверхностью. Микровинтами достигалось позиционирование, при этом дополнительно были задействованы вращающийся шарниром угловой позиционер и вертикальный угловой позиционер с винтовыми стопорами. В перспективе ведущихся работ ведется подготовка к полной механизации установки за счет использования системы шаговых двигателей, управляемых с GUI.

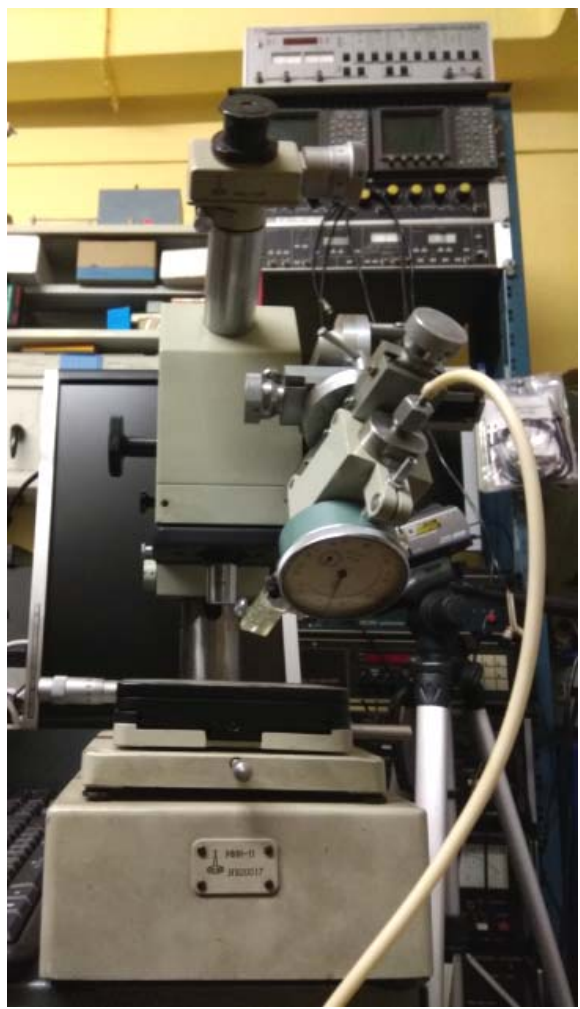

Рис. 1. Общая конфигурация установки.

Вся арматура инъектора, с целью уточненного позиционирования, привинчивалась винтами к корпусу интерферометра, за счет чего достигалась вибрационная согласованность измерений (вместо монтажа на столе лазерной установки или оптических рельсах / напольной оптической скамье). Координатным столом осуществлялось перемещение в частности – вращение в горизонтальной ротационной сетке с угловой разметкой в полярных координатах, носителей образцов, в частности чипов, в силу чего являлось возможным установление угловых зависимостей сигнализации клеток, их векторной / угловой направленности морфогенеза (нейрогониометрия на чипе) и ориентации в экспериментальных полях (см. рис. 2).

В качестве метрологической схемы измерений использовались фазочувствительные (т.е. синхронные по детектированию) схемы усиления и анализа (Princeton Applied Research 5208 – Two Phase Lock in Analyzer и Princeton Applied

Research 5210 – Dual Phase Lock-in Amplifier). Они видны на заднем плане на рис. 1, наряду с частотно-перестраиваемым фильтром модели KEMO VBF/8 (Dual Channel Variable Filter). Более подробно результаты измерений изложены в отдельной статье.

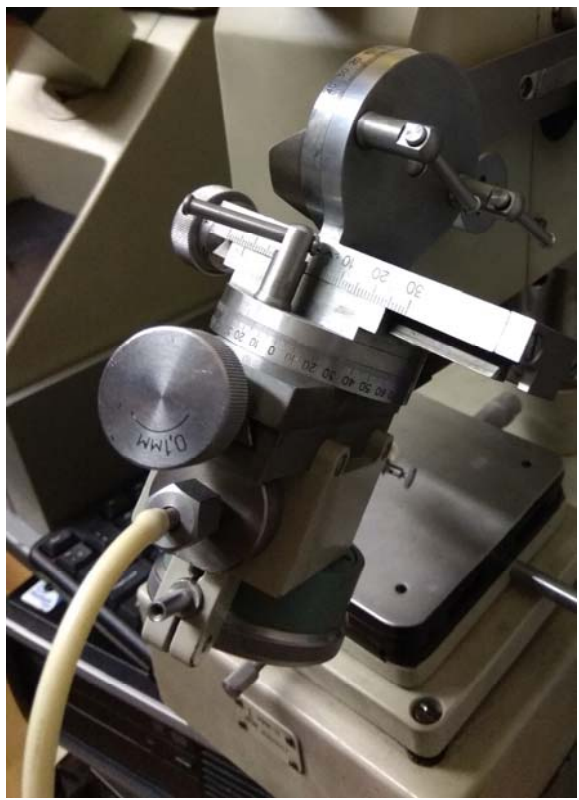

Рис. 2. Системы углового позиционирования микроманипулятора.

В настоящее время ведутся работы по адаптации интерференционного метода для иных способов микроинъекции в клеточные тест-системы не только для чипов биотестирования, но и для усеченных 96-луночных (и иных) планшет. В частности, готовится интерференционная система регистрации на альтернативных принципах для схем микроинъекции с капиллярной хроматографической микроколонки, как это показано на рис. 3. Данная система может также работать в режиме снятия электрического сигнала с корпуса колонки. Дополнительно может быть введен зонд непосредственно в планшетную лунку с помощью микроманипулятора типа «OPTON» с управлением перемещением джойстиком с пульта, подающим сигнал на шаговые двигатели (изображен на заднем плане рис. 3).

**Обсуждение**

Данная конструкция физически исходит «из первых принципов» капиллярной оптики и базирующихся на ней технических средств анализа. При расчетах принимаются во внимание многолучевая интерференция в капилляре и формирование трехмерных интерференционных паттернов в нём [87,88] (для нестеклянных микроэлектродов сверхмалого диаметра – таких как углеродные волоконные микроэлектроды – это метрологически существенно по физическим причинам, поскольку абляционные паттерны лазерной интерференции приводят к изменению уровня сигнала энзиматического детектирования [89]). Учитываются нелинейные зависимости показателя преломления среды [90] (в частности, для витальных красителей, которыми может быть окрашена клетка – которые могут поверхностно взаимодействовать с капилляром [91] и диффузионно изменять граничные параметры заполнения капилляра электролитом). Исходя из классических физико-математических представлений [92], делается возможным измерение капиллярных констант вязких (превалентно, неньютоновских) жидкостей, к которым отчасти относится содержимое экспериментальной платформы, содержащей клетку и внеклеточную среду. Для методов микрокоагуляции и микроабляции клеток (имплементируемых, начиная с работ Чахотина, в УФ-диапазоне, а начиная с XXI века – с использованием микрокапилляров как световодов / оптических волноводов для лазерного излучения, вводимого в клетку, либо с использованием капиллярных эмиссионных дуг) возможно рассчитать интерференционные и эмиссионные эффекты селективно к используемым длинам волн вводимого манипулятором в клетку излучения; аналогично обстоит дело для эмиссионного спектрального микроанализа и позиционно-чувствительных методов лазерной искровой спекроскопии тканей [93].

Наивысшее геометрическое и технико-методическое подобие методов анализа отклика, обусловленного вводом капилляров (в частности – электрофизиологического отклика), имеет комплекс интерферометрических подходов, используемых в капиллярном электрофорезе или капиллярном микроэлектрофорезе. Интерференция обратно отраженного пучка на капилляре-колонке является источником данных автометрии показателя преломления при капиллярном электрофорезе [94]. Детектирование показателя преломления при лазерной интерферометрии, как правило, лежит в основе техник лазерного детектирования в капиллярном электрофорезе, работающих на принципах рефрактометрии когерентного излучения [95]. Для эквивалентных методов интерферометрии сред (с использованием капилляров) необходим учет собственного показателя преломления стенок капилляра [96]. Подобные подходы лежат в основе измерений биологических агентов в капиллярно-интерференционных методах [97], в том числе – на базе флуоресцентных, а не обычных подходов; в частности, применение стационарных паттернов интерференции в капиллярном электро-

MORPHOLOGIA • 2017 • Том 11 • № 4 • МОРФОЛОГІЯ

форезе обнаруживает эффективность и для принципов детектирования, ассоциированных с флуоресцентной корреляционной спектроскопией [98]. В настоящее время методы, основанные на измерениях интерференции на обратно отраженном пучке в капиллярном электрофорезе получили ещё один нестандартный тренд развития – TLS – в сопряжении с термолинзовой спектроскопией (особо – коаксиальной TLS [99]), однако это практически бессмысленно имплементировать, интегрируя с микроскопией при эксперименте на живой ткани, поэтому данный тренд в настоящем обсуждении не рассматривается.

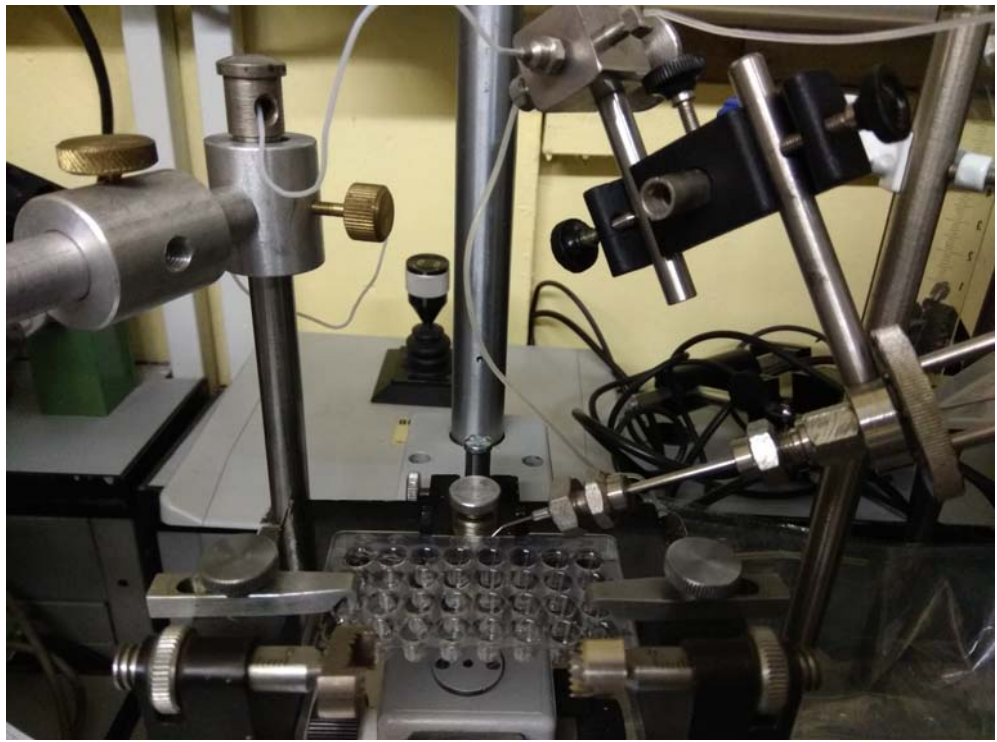

Рис. 3. Стенд монтажа интерференционной системы регистрации для схемы микроинъекции с капиллярной хроматографической микроколонки с параллельным снятием сигнала с колонки.

**Градов О.В., Насіров Ф.А., Скринник А.А., Яблоков А.Г. Просте методичне пристосування для мікроін'єкторних маніпуляцій і вимірювань на електроморфологічному чипі при мікроінтерферометричному контролі інтерфейсних і мембранних процесів на діапазоні товщини від 50 до 10000 ангстрем під різними кутами.**

**Реферат.** Мікроманіпуляції, перфузії і вимірювання, що проводяться з використанням скляних мікроелектродів, заповнених, як правило, електролітом, є класичною технікою експериментально-морфологічних і мембранно-електрофізіологічних досліджень на рівні окремих клітин і мембранних поверхонь. Стандартний (ефективний) діаметр скляного мікроелектрода в кінцевій області становить від 500 нм до менш ніж 100 нм, що перешкоджає використанню стандартних оптичних мікроскопів для його спостереження, відповідно до оптичних критеріїв (критерій Релєя і т.п.), оскільки при діаметрі конуса менше 500 нм він губиться в інтерференційної облямівці. Мікропроцесорним програмуванням пуллера (мікрокузні), що забезпечує витягування і розрив, хоча і можна досягти в відомих режимах заданих форм і діаметра кінця мікропіпеток, цей результат не є в повній мірі контрольованим в силу вищевказаних обмежень. У зв'язку з цим необхідне створення пристроїв контролю кінцевого фрагмента мікропіпеток як при отриманні, так і при експлуатації (внутрішньоклітинному або екстрацелюлярному введенні) в штатному режимі. При цьому необхідно, щоб даний метод дозволяв візуалізувати на зображенні клітини з мікроелектродами в реальному часі процеси, що відбуваються між ними, в залежності від типу і стану електрода, що дозволить нівелювати артефакти, з частотою систематичної помилки, що виникають при неконтрольованій експлуатації кінця мікропіпеток після застосування різних способів заливки електроліту (капілярного по Тасакі; вакуумного заповнення; заповнення спиртом з подальшим витісненням спирту по еквівалентній об'ємній характеристиці електролітом; заливка легкоплавкими сплавами як альтернатива рідким електролітам, що полегшує введення контакту хлорсрібного дроту). Нами пропонується конфігурація установки, що вирішує всі вищевказані проблеми шляхом введення інтерферометричного пристрою для мікроскопічного контролю мікроелектродів і мікроманіпулятора або мікроперфузора, вперше для даного типу оптичних приладів комбінованого з інтерферометричною оптичною схемою.

**Ключові слова:** мікроелектроди, мікроінтерферометр, мікроперфузія, патч-кламп, мікроманіпуляція, перфузія одиночних клітин, цитоелектрофізіологічний чіп.